\documentclass{article}
\usepackage[english]{babel}
\usepackage{amsmath,amsfonts,amsthm,amssymb,amscd}

\renewcommand{\Re}{\mathop{\rm Re}\nolimits}
\renewcommand{\Im}{\mathop{\rm Im}\nolimits}

\newcommand{\e}{\varepsilon}

\newcommand{\C}{{\mathbb C}}
\newcommand{\R}{{\mathbb R}}

\newcommand{\pP}{{\mathbb P}}
\newcommand{\I}{{\mathbb I}}
\newcommand{\E}{{\mathbb E}}

\newcommand{\N}{{\mathbb N}}
\newcommand{\la}{\lambda}

\newcommand{\La}{\Lambda}

\newcommand{\BB}{{\cal B}}
\newcommand{\CC}{{\cal C}}
\newcommand{\DD}{{\cal D}}

\newcommand{\FF}{{\cal F}}

\newcommand{\PP}{{\cal P}}
\newcommand{\RR}{{\cal R}}

\newcommand{\UU}{{\cal U}}

\newcommand{\lag}{\langle}
\newcommand{\rag}{\rangle}

\newcommand{\dd}{{\textup d}}

\newcommand{\PPPP}{{\mathfrak P}}

\newcommand{\BBBBB}{{\mathcal B}}

\newcommand{\dist}{\mathop{\rm dist}\nolimits}

\theoremstyle{plain}
\newtheorem{theorem}{Theorem}[section]
\newtheorem{lemma}[theorem]{Lemma}

\newtheorem{condition}[theorem]{Condition}
\theoremstyle{remark}
\newtheorem{remark}[theorem]{Remark}
\newtheorem{example}[theorem]{Example}
\newcommand{\de}{\delta}
\numberwithin{equation}{section}
\begin{document}
\author{Vahagn Nersesyan}
\date{}
\title{Exponential mixing for finite-dimensional approximations of the  Schr\"odinger equation with  multiplicative noise}
\date{}
 \maketitle
\begin{center}
 Laboratoire de Math\'ematiques,
Universit\'e de Paris-Sud XI\\ B\^atiment 425, 91405 Orsay Cedex,
France\\ E-mail: Vahagn.Nersesyan@math.u-psud.fr
\end{center}
{\small\textbf{Abstract.} We study the ergodicity of
finite-dimensional approximations of the Schr\"o\-din\-ger equation.
The system is driven by a multiplicative scalar noise. Under general
assumptions over the distribution of the noise,  we show that the
system has a unique stationary measure $\mu$ on the unit sphere $S$
in $\C^n$, and  $\mu$ is absolutely continuous with respect to the
Riemannian volume  on $S$. Moreover, for any initial condition in
$S$, the solution converges exponentially fast to the measure $\mu$
in the variational norm.}
 \tableofcontents

\section{Introduction}\label{S:Intro}
We consider the problem
\begin{align}
i\frac{\dd z}{\dd t} &=\La z+\beta(t) Bz+\e F(z),\label{E:hav}\\
z(0)&=z_0,\label{E:sp}
\end{align}
where $\La$ and  $B$ are
 Hermitian matrices, $F: \C^n\rightarrow \C^n$ is a real-analytic
 function such that the scalar product $\lag F(z),z\rag$ is real for any
 $z\in \C^n$, and $\e\in\R$ is a small constant. 
We  assume that $\beta(t)$ is a random process of the
 form
\begin{equation}\label{E:pu}
\beta (t)=\sum_{k=0}^{+\infty}I_{k}(t)\eta_k(t-k),\,\,\,\,\,t\ge 0,
\end{equation}
where $I_{k}(\cdot)$ is the indicator function of the interval
$[k,k+1)$ and $\eta_k$ are independent identically distributed
(i.i.d.) random variables in $L^2([0,1],\R)$.

The restriction of the solution of (\ref{E:hav}), (\ref{E:sp}) at
integer times formes a Markov chain. The aim of this paper is the
study of ergodicity of this chain. 
Noting that the unit sphere $S$ in $\C^n$ is invariant under the
flow defined by the equation, we show that the chain in question has
a unique stationary measure  on  $S$. Moreover, it is proved that
this measure is exponentially attracting in the variational norm.
Once we have the uniqueness of stationary measure on the sphere $S$,
using the invariance of $S$, the class of all stationary measures in
$\C^n$ can be described.


The ergodicity of finite-dimensional stochastic systems is studied
by many authors. Let us mention some earlier results in this
direction. Uniqueness of stationary measure for non-degenerate
diffusion processes is obtained by Hasminskii \cite{Has}. The case
of the degenerate diffusions  is considered by Arnold and Kliemann
\cite{AL} and Veretennikov \cite{VP,VER}. 
  Various sufficient conditions
for ergodicity of abstract Markov processes are obtained by  Meyn
and Tweedie in \cite{MT1} and \cite{MT2}. 
  E and  Mattingly \cite{EM} consider the
finite-dimensional approximations of the 2D Navier--Stokes
equations, and      Romito \cite{MR}  considers the approximations
of the 3D Navier--Stokes equations. In both cases, the perturbation
is an additive white noise and the main result is the exponential
mixing in the variational norm.




The main difference between  this paper   and the earlier results
dealing with stochastic differential equations is that the noise is
not supposed to have a Gaussian structure. We prove the ergodicity
of system (\ref{E:hav}), (\ref{E:pu}) under some conditions over the
matrices $\La$ and $B$ and over the distribution of the random
variable $\eta_1$. Roughly speaking, we assume that there is no
proper vector space invariant under both $\La$ and $B$, and that the
support of the law of $\eta_1$ contains a ball of sufficiently high
dimension. These conditions enable us to use a measure
transformation theorem from \cite{AKSS} and some controllability
results from \cite{BCH} and \cite{BCMR}.


Our proof is based on a classical coupling argument combined with
some controllability properties of the  Schr\"odinger equation. It
is divided into two steps. First, using the measure transformation
theorem, we show that there is a ball $D\subset S$ and a constant
$p\in(0,1)$ such that the variational distance at time $t=1$ between
any two solutions issued from $D$ is less than $p$. Then we show
that $D$ is accessible from any point of $S$. From the compactness
of $S$ it follows that the
 first hitting time of $D$ admits an exponential estimate. Combination
 of the above properties with a suitable coupling
construction gives the proof of the exponential mixing property.

Let us note that in the case of the diffusion process defined by the
Stratonovich  stochastic differential equation
$$
i{\dd z}=[\La z+\e F(z)]{\dd t}+ Bz\circ W(t),
$$
the uniqueness of stationary measure can be obtained as a
consequence of \cite{AL}. Indeed, under the conditions imposed on
$\La$ and $B$, a direct verification shows that the Lie algebra
generated by the drift and diffusion fields is full at the point
$e_1$, where $e_1$ is the first eigenvector of the matrix $\La$.

As an application of our result, we consider the Galerkin
approximations of the Schr\"odinger equation with potential of
random amplitude. We show that if the deterministic part of the
potential is in a general position, then   the property of
exponential mixing holds for any finite-dimensional approximation.
In conclusion, let us note that even though our proof does not apply
to the infinite-dimensional Schr\"odinger equation, many properties
remain valid. In particular,  an approximate
controllability property holds  for the Schr\"odinger equation, which
enables one  to show that almost any trajectory of randomly forced
  equation is unbounded in the Sobolev space of any order
$s>0$. These questions will be addressed in a forthcoming paper.
\\
\textbf{Acknowledgments.} The author is grateful to his
advisor, Armen Shirikyan, for many helpful conversations and
support.
\\



\textbf{Notation}
\\\\
In this paper, we use the following notation. \\$S$ is the unit
sphere in $\C^n$, i.e. $S=\{x\in \C^n: \|x\|_{\C^n}=1\}$. $S$ is
regarded as a $(2n-1)$-dimensional  real-analytic manifold endowed
with the standard Riemannian metric and the corresponding measure.
The latter is denoted by $m$.
\\ $T_y S$ is the tangent space to $S$
at the point $y\in S$, i.e. $T_y S=\{x\in \C^n: \Re\langle x,y\rangle=0\}$, where
$\langle\cdot,\cdot\rangle$ stands for the scalar product in $\C^n$.\\ $C_b(S)$
is the space of real-valued continuous bounded functions on $S$
endowed with the norm $\|f\|_\infty:=\sup|f|$.\\$\BB(S)$ is the
Borel $\sigma$-algebra of $S$.\\$\PP(S)$ is the set of probability
measures on $(S,\BB(S))$.
\\ The set $\PP(S)$ is endowed with the    variational norm:
$$
\|\mu_1-\mu_2\|_{var}:=\sup_{\Gamma\in\BB(S)}|\mu_1(\Gamma)-\mu_2(\Gamma)|,\,\,\,\,\,
\mu_1,\mu_2 \in \PP(S).
$$
The distribution of a random variable $\xi$ is denoted by
$\DD(\xi)$.
\\ The indicator function of  a set $\Gamma$ is denoted by
$I_{\Gamma}$.\\
For a metric space $E$, we denote by $B_E(a,r)$ the open ball of radius
$r>0$ centered at $a\in E$. If $E=\C^n$, we  simply write $B(a,r)$.\\
$\I$ denotes the set of irrational numbers.

\section{Main result}

\subsection{Uniqueness and exponential mixing}

Under the conditions described at the beginning of Section
\ref{S:Intro}, for any $z_0\in\C^n$ problem (\ref{E:hav}),
(\ref{E:sp}) has a unique solution almost surely belonging to the
space $C([0,\infty),\C^n)$. Let $\UU_t^\e:\C^n\rightarrow \C^n$ be
the resolving operator of (\ref{E:hav}), (\ref{E:sp}). Note that
\begin{equation}\label{E:barev}
\|\UU_t^\e(z_0)\|_{\C^n}=\|z_0\|_{\C^n},\,\,\,\,t\ge0.
\end{equation}
Let $z_0$ be a  $\C^n$-valued  random variable independent of
$\{\eta_k\}$. Denote by $\FF_k$  the $\sigma$-algebra generated by
$z_0,\eta_0,\dots,\eta_{k-1}$.  Then $\UU_k^\e(z_0)$ is a
homogeneous Markov chain with respect to $\FF_k$. The corresponding
transition function has the form
$P_k^\e(z,\Gamma)=\pP\{\UU_k^\e(z)\in\Gamma\}$, $z\in \C^n$,
$\Gamma\in \BBBBB (\C^n)$, and the Markov operator  is defined as
$$
\PPPP^{\e*}_k\mu(\Gamma)=\int_{\C^n}
P_k^\e(z,\Gamma)\mu(\dd z),
$$ where $\mu\in\PP(\C^n).$ Recall that a  measure $\mu\in\PP(\C^n)$ is called stationary for
(\ref{E:hav}), (\ref{E:pu}) if $\PPPP_1^{\e*}\mu=~\mu$.

It follows from (\ref{E:barev}) that the unit sphere $S$ is
invariant under the flow  defined by (\ref{E:hav}). The
Bogolyubov--Krylov argument and the compactness of $S$ imply the
existence of a stationary measure $\mu\in \PP(S)$ for problem
(\ref{E:hav}), (\ref{E:pu}).


 To be able to show the uniqueness of
stationary measure, we need the following conditions.
\begin{condition}\label{C:c}
The random variables $\eta_k$  have the form
$$
\eta_k(t)=\sum_{j=1}^\infty b_j\xi_{jk}g_j(t), \,\,\,\, t\in[0,1],
$$where $\{g_j\}$ is an orthonormal basis in $L^2([0,1],\R)$,
$b_j\ge0$ are constants with
$$
 \sum_{j=1}^\infty b_j^2<\infty,
$$ and $\xi_{jk}$ are independent real-valued random variables such that
$\E\xi_{jk}^2=1$. Moreover, the distribution of $\xi_{jk}$ possesses
a continuous density $\rho_j$ with respect to the Lebesgue measure
and $\rho_j(0)>0$.
\end{condition}This condition is adapted to the hypotheses of  a
measure transformation theorem from \cite{AKSS}. In particular,
under this condition,   the image of  measure $\DD(\eta_k)$ under a
large class of finite-dimensional transformations is absolutely
continuous with respect to the Lebesgue measure.

Let $\{e_j\}_{j=1}^n$ be the set of normalized eigenvectors of $\La$
with eigenvalues $\la_1\le\ldots\le\la_n$.
\begin{condition}\label{C:c2}
The eigenvalues of $\La$ are distinct and $\lag Be_1,e_j\rag\neq0$,
$j=~1,\dots,n$.
\end{condition}
Under this condition, some strong controllability properties hold
for (\ref{E:hav}). In particular, the linearization of (\ref{E:hav})
is controllable, which, combined with the inverse function theorem,
gives a local exact controllability property (see Section
\ref{S:CCC}). Moreover, Condition \ref{C:c2} also allows us to use a
stabilization result from \cite{BCMR}. Notice that, as in
\cite{BCMR}, all the results of the paper remain valid under the
assumption that for some $i=1,\ldots,n$ we have $\lag
Be_i,e_j\rag\neq0$ for all $j=~1,\dots,n$ and
$|\la_p-\la_i|\neq|\la_q-\la_i|$ for all $p\neq q$. Clearly, in the
case $i=1$, the last condition follows from the  non-degeneracy of
the spectrum of $\La$.

The following theorem is our main result.
\begin{theorem}\label{T:him}
 Suppose that Conditions \ref{C:c} and \ref{C:c2} are  satisfied. Then there is an integer $N\ge1$ and a constant $\e_0>0$ such that, if
\begin{equation}\label{E:pp}
b_j\neq0,\,\,\,\, j=1,\dots,N,
\end{equation} then problem
 (\ref{E:hav}), (\ref{E:pu}) has a unique stationary measure $\mu \in
 \PP(S)$ for  $|\e|<\e_0$. Moreover, $\mu$ is absolutely continuous with respect to the  measure $m$, and for any initial  measure $\nu\in\PP(S)$, we have
\begin{equation}\label{E:hima}
\|\PPPP_k^{\e*}\nu-\mu\|_{var}\le C e^{-c k}, \,\,\,\, \text{ $k\ge1$},
\end{equation}where $C>0$ and $c>0$ are constants.
\end{theorem}

\subsection{Proof of Theorem \ref{T:him}} The proof of Theorem
\ref{T:him} is derived from the two lemmas below. Their proofs are
given in Section \ref{S:le1}.
\begin{lemma}\label{L:xt}
Under the conditions of Theorem \ref{T:him}, there are constants
$\delta_0>0$, $\e_0>0$ and $p\in (0,1)$ and   integers $N\ge 1$
and $l\ge1$ such that, if (\ref{E:pp}) holds, then:
\begin{itemize}
\item[(i)]For any $z, z'\in S\cap
B(e_1,\delta_0)$ and $|\e|<\e_0$, we have
\begin{equation}
\|P_1^\e(z,\cdot)-P_1^\e(z',\cdot)\|_{var}\le p\nonumber.
\end{equation}
\item[(ii)] For any $z\in S$ and $|\e|<\e_0$, the measure 
 $P_l^\e(z,\cdot)$ is absolutely continuous with respect to $m$.
\end{itemize}
\end{lemma}
For any $\delta>0$, let us introduce the stopping time
\begin{equation}
\tau_{\delta,\e}=\min\{k\ge0:\UU_k^\e(z)\in
B(e_1,\delta)\}.\nonumber
\end{equation}
\begin{lemma}\label{L:tau}
Under the conditions of Theorem \ref{T:him}, for any $\delta>0$ there is a constant
$\e_\delta>0$ and an integer $N\ge1$ such that, if (\ref{E:pp})
holds, then
\begin{equation}\label{E:tau}
\E_{z} e^{\alpha \tau_{\delta,\e} }\le C\,\,\,\,\,\,\,\,\, \text{for
all $z\in S$ and $|\e|<\e_\delta$,} \end{equation} where $\alpha>0$
and $C>0$ are constants, and the subscript $z$ means that the
expectation is taken for the chain issued from $z$.
\end{lemma}

\vspace{6pt}
\begin{proof}[Proof of Theorem \ref{T:him}]
\textbf{Step 1.} Let $z_0, z_0'\in S$. The idea of the proof is to
construct two sequences $y_k$ and $y_k'$ such that
$\DD(y_k)=P_k^\e(z_0,\cdot)$, $\DD(y_k')=P_k^\e(z_0',\cdot)$ and the
following inequality holds
\begin{equation}\label{E:bav}
\|\DD(y_k)-\DD(y_k')\|_{var}\le C e^{-c k}, \,\,\,\, \text{
$k\ge0$}.
\end{equation}A well-known argument shows that  (\ref{E:bav}) implies
(\ref{E:hima}) (e.g, see \cite{KS1}).

\vspace{6pt} \textbf{Step 2.} Let $z, z'\in S$. If $z=z'$, then
define $V(z,z')=V'(z,z')=\UU_1^\e(z)$. Let $\delta_0>0$ and $\e_0>0$
be the constants in Lemma \ref{L:xt} and $|\e|<\e_0$. If $z\neq z'$
and $z,z'~\in B(e_1,\delta_0)$, then let $V(z,z')$, $V'(z,z')$ be
any maximal coupling for $(P_1^\e(z,\cdot),  P_1^\e(z',\cdot))$ (see
\cite{Li}, Section I.5). Otherwise, let $V(z,z')$ and $V'(z,z')$ be
the values at $t=1$ of the solutions of the following problems:
\begin{displaymath}
\begin{array}{ll}
i\frac{\dd y}{\dd t}=\La y+\eta(t) By+\e F(y), & i\frac{\dd y'}{\dd t} =\La y'+\eta'(t) By'+\e F(y'),\\
y(0)=z, & y'(0)=z',\\
\end{array}
\end{displaymath}
where $\eta$ and $\eta'$ are independent random variables with
$\DD(\eta)=\DD(\eta')=\DD(\eta_1)$. Let $V_k,V_k',$ $k\ge1$ be
independent copies of the random variables $V$ and $V'$ depending on
the parameters $z$ and $z'$. Let $y_0=z_0$ and $y_0'=z_0'$. Define
$y_k$ and $y_k'$, $k\ge 1$ by the relations
 \begin{align}
y_k&=V_k(y_{k-1},y_{k-1}'), \nonumber\\
y_k'&=V_k'(y_{k-1},y_{k-1}').\nonumber
\end{align}
Clearly, $(y_k, y_k')$ is a Markov chain. It is easy to see that
$\DD(y_k)=P_k^\e(z_0,\cdot)$ and $\DD(y_k')=P_k^\e(z_0',\cdot)$.
Define
\begin{equation}\label{E:nort}
T=\min\{k\ge0:y_k,y_k'\in B(e_1,\delta_0)\}.
\end{equation}Using the same arguments as in the proof of
(\ref{E:tau}), one can show that, if $\e_0>0$ is sufficiently small,
then
\begin{equation}\label{E:Tau}
\E e^{\alpha T}\le C\,\,\,\text{for}\,\,\,|\e|<\e_0,
 \end{equation}
where  $\alpha>0$ and $C>0$ are some constants not depending on $\e$
(see Remark \ref{R:norr}).

\vspace{6pt} \textbf{Step 3.} Suppose that there is a random integer
$\ell$ such that
\begin{align}
&y_k=y_k'\,\,\,\,\, \text{for all $k\ge\ell$},\label{E:LL1}\\
&\E e^{\gamma \ell}\le C.\label{E:LL2}
\end{align} Then (\ref{E:LL1}) and (\ref{E:LL2})  imply (\ref{E:bav}). Indeed, for any $f\in
C_b(S)$, $\|f\|_\infty\le 1$, we have
\begin{align}\label{E:vn}
\big|\E(f(y_k)-f(y_k'))\big|&\le\E \big|(f(y_k)-f(y_k'))\big|
\nonumber\\&\le\E\big|(f(y_k)-f(y_k')) I_{\{k\ge
  \ell\}}\big|+\E\big|(f(y_k)-f(y_k')) I_{\{k<
  \ell\}}\big|\nonumber\\&\le2 \pP\{k<\ell\}\le 2C
e^{-\gamma k},\nonumber
\end{align}which proves (\ref{E:bav}).

\vspace{6pt} \textbf{Step 4.} Let us introduce the stopping times
$T(0)=0$, $T(1)=T$ and
$$
T(n)=\min\{k>T(n-1):y_k,y_k'\in B(e_1,\delta_0)\},\,\,\,\,n\ge2.
$$
Using the strong Markov property and (\ref{E:Tau}), we see that
$$
\E e^{\alpha T(n)}=\E e^{\alpha
T(n-1)}\E_{Y(n)}
e^{\alpha T}\le C\E e^{\alpha T(n-1)},
$$ where $Y(n)=(y_{T({n-1})},y_{T({n-1})}')$.
Thus
\begin{equation}\label{E:EE}
\E e^{\alpha T(n)}\le C^n.
\end{equation}
Define
$$
\ell=\min\{k\ge0: y_n=y_n'\,\,\,\, \text{for all $n\ge k$}\},
$$ where $\min\{\emptyset\} =\infty$.
Let us show that
\begin{equation}\label{E:pen}
\pP\{\ell>T(n+1)\}\le p^n,
\end{equation} where $p\in (0,1)$ is the constant in Lemma \ref{L:xt}. Indeed, it follows from Lemma
\ref{L:xt} and  the construction of $y_k$ and $y_k'$ that
\begin{align}
\pP\{y_{T(n)+1}\neq y_{T(n)+1}'\}&=\pP\{y_{T(n)+1}\neq
y_{T(n)+1}'|y_{T(n)}\neq y_{T(n)}'\}\pP\{y_{T(n)}\neq
y_{T(n)}'\}\nonumber\\&\le p \pP\{y_{T(n)}\neq
y_{T(n)}'\}\le p\pP\{y_{T(n-1)+1}\neq y_{T(n-1)+1}'\}.\nonumber
\end{align}Iteration of this inequality gives
$$
\pP\{y_{T(n)+1}\neq y_{T(n)+1}'\}\le p^n.
$$On the other hand, the definition of $\ell$  implies that
$$
\pP\{\ell>T(n+1)\}\le\pP\{y_{T(n)+1}\neq y_{T(n)+1}'\},
$$which proves (\ref{E:pen}).
Thus, by the Borel--Cantelli lemma, we have $\pP\{\ell<\infty\}=1$.

Let $r>0$ be so large that $C^\frac{1}{r}p^{1-\frac{1}{r}}<1$, and
let $c$ be so small that $r c<\alpha$. Using the Cauchy--Schwarz
inequality and (\ref{E:pen}), we obtain
\begin{align}
\E e^{c \ell}&\le1+ \sum_{n=0}^\infty \E (I_{\{T(n)<\ell\le
T({n+1})\}} e^{c \ell})\nonumber\\&\le1+\sum_{n=0}^\infty \E
(I_{\{T(n)<\ell\le T({n+1})\}} e^{c T({n+1})})\nonumber\\&\le
1+\sum_{n=0}^\infty (\E e^{rc
T({n+1})})^{\frac{1}{r}}\pP\{\ell>T(n)\}^{1-\frac{1}{r}}\nonumber\\&\le1+C^\frac{1}{r}p^{\frac{1}{r}-1}\sum_{n=0}^\infty
(C^\frac{1}{r}p^{1-\frac{1}{r}})^n<\infty.\nonumber
\end{align}
This completes the proof of (\ref{E:hima}).

\vspace{6pt} \textbf{Step 5.} To show that the stationary measure
$\mu$ is absolutely continuous with respect to the Riemannian volume
$m$ on $S$, take any $\Gamma\in\BB(S)$ such that $m(\Gamma)=0$
and let $l\ge1$ be the integer in Lemma \ref{L:xt}. Then
$$
\mu(\Gamma)=\int_S P_l^\e(z,\Gamma) \mu(\dd z)=0,
$$ as $P_l^\e(z,\cdot)$ is absolutely continuous with respect to $m$ for any $z\in S$.
\end{proof}
\subsection{Stationary measures in $\C^n$}
 Any measure $\nu\in\PP(\C^n)$ can be written in the form
\begin{equation}\label{E:opi}
\nu=\alpha\de_0+(1-\alpha)\bar{\nu},
\end{equation}where $\alpha\in[0,1]$, $\de_0$ is the Dirac measure concentrated at
zero and $\bar\nu\in\PP(\C^n\backslash\{0\})$. Indeed, it suffices
to take $\alpha=\nu(\{0\})$ and
$\bar\nu(\cdot)=\frac{1}{1-\alpha}\nu(\cdot\cap\C^n\backslash\{0\})$,
if $\alpha<1$. On the other hand, for any measure
$\bar\nu\in\PP(\C^n\backslash\{0\})$ there is a measure
$\gamma\in\PP(\R_+^*)$ and a random measure $\mu_r\in\PP(S)$, $r\in
\R_+^*$ (i.e. for any $\Gamma\in \BB(S)$ the function
$r\rightarrow\mu_r(\Gamma)$ is measurable) such that for any bounded
measurable function $f:\C^n\backslash\{0\}\rightarrow\R$ we have
\begin{equation}\label{E:uy}
\int _{\C^n\backslash\{0\}} f(v) \bar\nu(\dd v)=\int _{\R_+^*}\int _{S} f(ru) \mu_r(\dd
u)\gamma(\dd r)\end{equation} (e.g., see \cite{AR}). In this case, we write
\begin{equation}\label{E:opt}
\bar\nu(\dd r,\dd
u)=\mu_r(\dd u)\gamma(\dd r).
\end{equation}

Let $\mu\in\PP(S)$ be the stationary measure in Theorem \ref{T:him}
for $\e=0$, i.e. corresponding to the linear equation.
\begin{theorem}\label{C:a}
Under the conditions of Theorem \ref{T:him},
 there is an integer $N\ge1$ such that, if
 (\ref{E:pp}) holds, then a measure $\nu\in\PP(\C^n)$
 is stationary for problem
 (\ref{E:hav}), (\ref{E:pu}) with $\e=0$ if and only if it can be represented in
 the form (\ref{E:opi}), (\ref{E:opt}) in a way that $\mu_r=\mu$
 for $\gamma$-almost all $r\in\R_+^*$. Moreover, for any initial  measure
 $\nu'\in\PP(\C^n)$ of the form
\begin{align}
\nu'&=\alpha\de_0+(1-\alpha)\bar{\nu}',\nonumber\\
\bar{\nu}'(\dd r,\dd
u)&=\mu_r'(\dd u)\gamma(\dd r),\nonumber
\end{align}
 we have
\begin{equation}\label{E:himai}
\|\PPPP_k^{0*}\nu'-\nu\|_{var}\le C e^{-c k}, \,\,\,\, \text{ $k\ge1$},
\end{equation}where $C>0$ and $c>0$ are constants.
\end{theorem}
\begin{proof}
Without loss of generality, we can assume that $\alpha=0$, i.e.
$\nu=\bar\nu\in\PP(\C^n\backslash\{0\})$. Suppose that $\nu$ is a
stationary measure. Let us show that $\mu_r$ is stationary for
$\gamma$-almost all $r\in\R_+^*$. Take any bounded measurable
functions $f:S\rightarrow\R$ and $g:\R_+^*\rightarrow\R$. By
(\ref{E:uy}), we have
\begin{align}
\int _{\C^n\backslash\{0\}} (fg)(v) \nu(\dd v)&=\int
_{\R_+^*}g(r)\int _{S}f(u) \mu_r(\dd u)\gamma(\dd
r)\nonumber\\&=\E\int _{\C^n\backslash\{0\}} (fg)(\UU_1^0(v))
\nu(\dd v)\nonumber\\&=\E\int
_{\C^n\backslash\{0\}}f\Big(\frac{\UU_1^0(v)}{\|\UU_1^0(v)\|}\Big)g(\|\UU_1^0(v)\|)
\nu(\dd v)\nonumber\\&=\int _{\R_+^*}g(r)\Big[\E\int
_{S}f(\UU_1^0(u)) \mu_r(\dd u)\Big]\gamma(\dd r)\label{E:tyr},
\end{align} where we used the fact that $\UU_t^0(v)$ is a solution of
a linear equation. As (\ref{E:tyr}) holds for any bounded measurable
functions $g$ and $f$ and the Borel $\sigma$-algebras on $S$ and
$\R_+^*$ are countably generated, we see that $\mu_r$ is stationary
for $\gamma$-almost all $r\in\R_+^*$. By the uniqueness of
stationary measure, we get $\mu_r=\mu$
 for $\gamma$-almost all $r\in\R_+^*$.

 On the other hand, if  $\mu_r=\mu$
 for $\gamma$-almost all $r\in\R_+^*$ and $\nu$ is defined by (\ref{E:opt}), then  by a similar argument, one
 can prove that $\nu$ is a stationary measure.

To prove the second assertion, let us  take any  bounded
measurable functions $f:S\rightarrow\R$ and $g:\R_+^*\rightarrow\R$
such that $\sup |f|\le1$ and $\sup |g|\le1$. Then
\begin{align}
&\E\Big|\int_{\C^n\backslash\{0\}}(fg)(\UU_k^0(v))\nu'(\dd
v)-\int_{\C^n\backslash\{0\}}(fg)(v)\nu(\dd
v)\Big|\nonumber\\&\le\E\int _{\R_+^*}|g(r)|\bigg|\int
_{S}f(\UU_k^0(u))) \mu_r(\dd u)-\int_{S}f(u)\mu(\dd
u)\bigg|\gamma(\dd r)\nonumber\\&\le C e^{-c k}\nonumber.
 \end{align}The general case is obtained by the monotone class theorem.
\end{proof}
\begin{remark} Denote by $S_r$   the sphere  of radius $r>0$ in $ \C^n$.
Let $N(r)\ge1$ and $\e_0(r)>0$  be the constants in Theorem
\ref{T:him} applied for the sphere $S_r$, and let $\mu_r\in\PP(S_r)$
be the stationary measure for $\e\in(0,\e_0(r))$. In this case, the
projections of measures $\mu_r$ to $S$ depend on $r>0$. On the other
hand, $N(r)\rightarrow\infty$ and $\e_0(r)\rightarrow0$ as
$r\rightarrow\infty$. Taking into account these facts, one can
reformulate Theorem \ref{C:a} for measures $\nu\in\PP(B(0,r))$.

\end{remark}
\section{Controllability results}\label{S:CCC}
 Let us consider the control system
\begin{align}
i\frac{\dd z}{\dd t} &=\La z+u(t) Bz,\label{E:hav2}\\
z(0)&=z_0,\label{E:sp2}
\end{align}where the state is $z$ and the control is $u$. Let
$\RR_t(\cdot,u):S\rightarrow S$ be the resolving operator of $(\ref{E:hav2}), (\ref{E:sp2})$. Recall that $\La$ and $B$ satisfy Condition \ref{C:c2}.
\begin{theorem}\label{T:Con}
System (\ref{E:hav2}), (\ref{E:sp2}) is globally exactly
controllable, i.e. for any $z_1, z_2\in S$ there is a time $T\ge0$
and a control $u\in L^2([0,T],\R)$ such that $\RR_T(z_1,u)=~z_2$.
\end{theorem}

The proof of this theorem can be derived from Theorem 4 in
\cite{AD}, where a necessary and sufficient condition for the
controllability of (\ref{E:hav2}) is stated. One can verify that the
condition given in \cite{AD} is weaker than Condition \ref{C:c2}.
Here we give another proof of Theorem \ref{T:Con}. This proof
provides some additional information on the control (see Remark
\ref{R:rem}), which is important for the application in the proof of
Theorem \ref{T:him}.

The proof of Theorem \ref{T:Con} is based on several ideas of \cite{BCH} and
\cite{BCMR}. It  is derived from two lemmas below, which are
of independent interest. We postponed the proof of Theorem \ref{T:Con} to the end of this section.

Let us introduce the following $(2n-1)$-dimensional subspace of
$L^2([0,1],\R)$:
$$
E_n=\{v\in L^2([0,1]): v(t)=\sum_{k=-(n-1)}^{n-1}d_ke^{i\mu_k
t},\,\,d_k\in \C,\, d_{-k}=\bar{d}_{k},\, t\in [0,1]\},
$$where  $\mu_{k}=\la_{k+1}-\la_1$ and $\mu _{-k}=-\mu_k$.
\begin{lemma}\label{L:1}
For any $\nu>0$ there is a constant $\delta>0$ such that  for any
$z_i\in S\cap B(e_1,\delta)$ and $z_f\in S\cap B(e_1 e^{-i\la_1
},\delta)$ there is a control $u\in B_{E_n}(0,\nu)$ satisfying
$\RR_1(z_i,u)=z_f.$
\end{lemma}
\begin{proof}
We follow the ideas of \cite{BCH}, where the local exact
controllability of an infinite-dimensional Schr\"odinger equation is
proved using the Nash--Moser implicit function theorem. In our
situation, the controllability is derived from the inverse function
theorem.

For any $z\in S$ and $u\in E_n$, define $\Phi(z,u)=(z,\RR_1(z,u))$.
Note that $\Phi(e_1,0)=(e_1,e_1e^{-i\la_1})$. We are going to show
that the conditions of inverse mapping theorem are satisfied in a
neighborhood of the point $(e_1,0)\in S\times E_n$. Clearly, $\Phi$
is continuously differentiable. Let  us show that mapping  $D\Phi
(e_1, 0):T_{e_1}S\times E_n\rightarrow T_{e_1}S\times
T_{e_1e^{-i\la_1 }}S$ is an isomorphism. Consider the linearization
of (\ref{E:hav2}), (\ref{E:sp2}) around $(e_1e^{-i \la_1 t},0)$:
\begin{align}
i\frac{\dd y}{\dd t} &=\La y+w(t) Be_1e^{-i\la_1  t } ,\label{E:hav3}\\
y(0)&=y_0,\label{E:sp3}
\end{align}
where $w\in E_n$ and $y_0\in T_{e_1}S$. Denote by $y_t=y_t(y_0,w)$
the solution of problem (\ref{E:hav3}), (\ref{E:sp3}).
 One can verify that
$D\Phi(e_1,0)(y_0,w)=(y_0,y_1)$. 
Note that (\ref{E:hav3}), (\ref{E:sp3}) is equivalent to
\begin{equation}\label{E:D}
y_t=e^{-i\La t}y_0-i\int_0^t e^{-i\La(t-s)}w(s)Be_1e^{-i\la_1 s }\dd
s.
\end{equation} Let $B_{ij}=\lag Be_i,e_j\rag$, $i,j=1,\dots,n$. Taking the scalar product of (\ref{E:D}) with
$e_ke^{-i\la_k}$, we obtain for $t=1$
\begin{equation}\label{E:DD}
\lag y_1,e_ke^{-i\la_k}\rag=\lag y_0,e_k\rag-iB_{1k}\int_0^1 e^{i \mu_{k-1}s}
w(s)\dd s.
\end{equation}Clearly $y_1(y_0,w)\in T_{e_1e^{-i\la_1}}S$, if $y_0\in T_{e_1}S$. Let $y_1'\in
T_{e_1e^{-i\la_1}}S$. By Condition~\ref{C:c2}, we have
$B_{1k}\neq0$. Hence, the equality $y_1(y_0,w)=y_1'$ is equivalent
to
\begin{equation}\label{E:s}
c_k :=\frac{\lag y_1',e_ke^{-i\la_k}\rag-\lag
y_0,e_k\rag}{-iB_{1k}}=\int_0^1 e^{i \mu_{k-1}s} w(s)\dd s,\,\,\,
k=1,\dots,n.
\end{equation}
Since  $(y_0,y_1')\in T_{e_1}S\times T_{e_1e^{-i\la_1 }}S$ and
$B_{11}\in \R$, we have $c_1\in \R$. As the functions
$\{e^{i\mu_ks}\}_{k=-(n-1)}^{n-1}$ are linearly independent, there
is a unique solution $w\in span \{e^{i\mu_ks}\}_{k=-(n-1)}^{n-1}$ of
the problem
\begin{align}
c_k=\int_0^1 e^{i \mu_{k-1}s} w(s)\dd s,\,\,\,\, \bar c_k=\int_0^1
e^{-i \mu_{k-1}s} w(s)\dd s,\,\,\,k=1,\dots,n.\nonumber
\end{align}Then $w=\bar w$, as $\bar w$ is a solution of the same
problem. Thus $w\in E_n$. This shows the surjectivity of $D\Phi
(e_1, 0)$. 
Finally, applying the inverse mapping
theorem, we conclude that $\Phi$ is a $C^1$ diffeomorphism in the
neighborhood of $(e_1,0)$.
\end{proof}
For any $u\in L^2([0,l])$, $l\in \N$, define $u_j \in L^2([0,1])$ as follows:
$$
u_j=u(j+\cdot)\big|_{[0,1]},\,\,\,\,j=0,\dots,l-1.
$$
\begin{lemma}\label{L:dk}
For any $\nu>0$, $\delta>0$ and $s\in \R$ the following assertions
hold.
\begin{itemize}
\item[(i)] For any $z_0\in S$ there
is a time $l\in \N$ and a control $u\in L^2([0,l],\R)$ such that
$\RR_l(z_0,u)\in S\cap B(e_1e^{is},\delta)$.
\item[(ii)] There is an integer $N\ge 1$ such that for any $z_0\in S$ the control $u$
in (i) can be chosen in a way that
\begin{equation}\label{E:norh}
u_j\in span\{g_1,\dots,g_N\}\,\,\,\,\,   \text{and}\,\,\,\,\,
\|u_j\|_{L^2([0,1])}\le~\nu, \,\,\,\,j=0,\dots,l-1  \end{equation}
\end{itemize}
\end{lemma}
\begin{proof}
\textbf{Step 1.} To prove (i), note that, without loss of generality, we can
assume that the first eigenvalue of $\La$ is of the form $\la_1=2\pi
\alpha$, where $\alpha\in \I$.

 Indeed, for any $\gamma\in \R$, define the matrix
$\La_\gamma:=\La+\gamma B$. Clearly, $\La_\gamma$ is an  Hermitian
matrix and Condition \ref{C:c2} is satisfied, if $|\gamma|$ is
sufficiently small. Let $\{\la_{k,\gamma}\}$ and $\{e_{k,\gamma}\}$
be the sets of eigenvalues and normalized eigenvectors of
$\La_\gamma.$ Clearly, the resolving operator of problem
(\ref{E:hav2}), (\ref{E:sp2}) with $\La$ replaced by $\La_{\gamma}$
is $\RR_\cdot(\cdot,\cdot+\gamma)$. First let us show that it is
possible to choose a sequence $\gamma_n\rightarrow 0$ such that
$\la_{1,\gamma_n}=2\pi\alpha_n$, where $\alpha_n\in\I$. Indeed,
suppose that for some $\eta>0$ we have $\la_{1,\gamma}=\la_1$ for
any $\gamma\in(-\eta,\eta)$. Then $\det(\La_\gamma-\la_1 I)=0$ for
any $\gamma\in(-\eta,\eta)$, where $I$ is the $n\times n$  identity
matrix. But $\det(\La_\gamma-\la_1 I)$ is a polynomial in $\gamma$,
and the coefficient of the first order term is $\lag
Be_1,e_1\rag(\la_2-\la_1)\cdot\ldots\cdot(\la_n-\la_1)$, which is
not zero by Condition \ref{C:c2}. This contradiction shows that
above-mentioned choice of the sequence $\gamma_n$ is possible.

 If (i) holds
for problem (\ref{E:hav2}), (\ref{E:sp2}) with $\La$ replaced by
$\La_{\gamma_n}$, then there are sequences $l_n\in \N$ and $u_n\in
L^2([0,l_n],\R)$ such that $\RR_{l_n}(z_0,\gamma_n+u_n)\in S\cap
B(e_{1,\gamma_n}e^{is},\frac{\delta}{2})$. If $n$ is sufficiently
large, we have $e_{1,\gamma_n}\in B(e_1,\frac{\delta}{2})$, thus
$\gamma_n+u_n$ is the desired control.


 Thus, we can suppose  that  $\la_1=2\pi
\alpha$, where $\alpha \in \I$. It follows that the set
$\{e^{-i\la_1k}: k\in\N\}$ is dense in the circle
$\{z\in\C:|z|=1\}.$

\vspace{6pt} \textbf{Step 2.} Here we prove that (i) holds, if
$\la_1=2\pi \alpha$, $\alpha\in \I$. Define $\CC=\{e_1e^{it}:t\in
\R\}$. It suffices to show that
\begin{itemize}
\item[(a)] For any $z_0\in S$ there
is a time $l\in \N$ and a control $u\in L^2([0,l],\R)$ such that
$\RR_l(z_0,u)\in \CC_\delta:=\{y\in S: \dist(y,\CC)\le\delta\}$.
\item[(b)] For any $z_0\in \CC_\delta$ there
is a time $k\in \N$ and a control $v\in L^2([0,k],\R)$ such that
$\RR_k(z_0,v)\in S\cap B(e_1e^{is},\delta)$.
\end{itemize}
 To prove (a), following the ideas of
\cite{BCMR}, we introduce the feedback design
\begin{equation}\label{E:con}
u(z)=c\Im (\lag Bz,e_1\rag\lag e_1,z\rag), \end{equation} where
$c>0$ is a small constant. Let us consider the problem
\begin{align}
i\frac{\dd z}{\dd t} &=\La z+u(z) Bz,\nonumber\\
z(0)&=z_0.\nonumber
\end{align}
As in \cite{BCMR}, one can show that for any $z_0\in S$ such that
$\lag z_0,e_1\rag\neq 0$, we have
$$
\lim_{t\rightarrow\infty}\dist(\RR_t(z_0,u(z(t))),\CC)= 0.
$$
The proof of our case is easier, as the linearization of
(\ref{E:hav2}), (\ref{E:sp2}) around the trajectory $(e_1,u\equiv0)$
is controllable. Now (a) follows from the fact that the set
$\{z_0\in S:\lag z_0,e_1\rag\neq 0\}$ is dense in $S$ and the
distance between two solutions of (\ref{E:hav2}) corresponding to
the same
control is constant.\\
The proof of (b) follows from the fact that   $\la_1=2\pi
\alpha$, $\alpha\in \I$. Indeed, consider the solution of
(\ref{E:hav2}), (\ref{E:sp2}) with control $u\equiv0$, that is
$e^{-i\La t}z_0$.
As $z_0\in\CC_\delta$, there is a $\tau\in \R$ such that $\|z_0-e_1e^{i\tau}\|_{\C^n}< \delta$.
Clearly, the set
$\{e_1e^{-i\la_1k+i\tau}:k\ge1\}$ is dense in $\CC$ and
 $$\|e^{-i\La
k}z_0-e_1e^{-i\la_1k+i\tau}\|_{\C^n}=\|z_0-e_1e^{i\tau}\|_{\C^n}<\delta.$$
Thus $e^{-i\La k}z_0\in B(e_1e^{is},\delta)$ for some $k\in \N.$
This completes the proof of (i).

\vspace{6pt} \textbf{Step 3.} To prove (ii), note that any $z_0\in
S$ has a neighborhood whose points are controlled to $S\cap
B(e_1e^{is},\delta)$ with controls from $span\{g_1,\dots,g_N\}$ for
some $N:=N(z_0)$. As $S$ is compact, we can find a universal
constant $N$. The second part of (ii) follows directly from the
construction (note that we can choose the constant $c$ in
(\ref{E:con}) arbitrarily small).
\end{proof}
\vspace{6pt}
\begin{proof}[Proof of Theorem \ref{T:Con}]
\textbf{Step 1.} Take any $z_1, z_2 \in S$. Thank to Lemma
\ref{L:1}, it suffices to find controls $u_j\in L^2([0,T_j],\R)$,
$T_j>0$, $j=1,2$ and a point $y\in S\cap B(e_1e^{-i\la_1},\delta)$
such that $\RR_{T_1}(z_1,u_1)\in S\cap B(e_1,\delta)$ and
$\RR_{T_2}(y,u_2)=z_2$.

\vspace{6pt} \textbf{Step 2.} Clearly, we can take as $u_1$ the
control provided by Lemma \ref{L:dk} for $z_0=z_1$ and $s=0$.

To construct the control $u_2$ and the point  $y$, first note that,
if $\RR_T(\bar z,u)=\bar y$, then $\RR_T( y, u')=z$, where
$u'(\cdot)=u(T-\cdot)$. Now let  $u\in L^2([0,T_2],\R)$ be the
control provided by Lemma \ref{L:dk} for $z_0=\bar{z}_2$ and
$s=\la_1$. It remains to take $u_2(\cdot)=u(T_2-~\cdot)$ and
$y=\overline{ \RR_T(\bar
  {z}_2,u)}$. Clearly, $y\in S\cap B(e_1e^{-i\la_1},\delta)$ and  $\RR_{T_2}(y,u_2)=z_2$.
       \end{proof}
\begin{remark}\label{R:rem}
It follows from the proof of Theorem \ref{T:Con} that for any
$\nu>0$
 there is an integer $N\ge1$ such that
 the control $u$ and the time $T$ can be chosen in a way  that $T\in
 \N$ and (\ref{E:norh}) is verified.
\end{remark}
Now let us consider the system
\begin{align}
i\frac{\dd z}{\dd t} &=\La z+u(t) Bz+\e F(z),\label{E:hav4}\\
z(0)&=z_0.\label{E:sp4}
\end{align} Let
$\RR_t^\e(\cdot,u):S\rightarrow S$ be the resolving operator of
$(\ref{E:hav4}), (\ref{E:sp4})$. Clearly, $\RR_t^0~\equiv~\RR_t$.
\begin{theorem}\label{L:taza}
For any $\nu>0$ and $\delta>0$  there is a constant
$\e_0=\e_0(\nu,\delta)>0$ and integers $N\ge1$ and $k\ge1$ such that
following assertions hold.
\begin{itemize}
\item[(i)] For any $z\in S$ there
is a control $u\in L^2([0,k],\R)$ such that $\RR_k^\e(z,u)\in
B(e_1,\delta)$ for all $|\e|<\e_0$.
\item[(ii)] For any $z\in S$ the control $u$
in (i) can be chosen  such that (\ref{E:norh}) is verified.
\end{itemize}
\end{theorem} Notice that the time $k\ge1$  is the same for all $z\in  S$.
\begin{proof}
\vspace{6pt} \textbf{Step 1.} First we prove (i) for $\e=0$.
Let us choose points $x_j\in S$, $j=1,\dots,q$ such that
$S\subset\cup_{j=1}^q B(x_j,\frac{\delta}{2})$ and $\lag
x_j,e_1\rag\neq0$. Using the arguments in Step 1 of the proof of
Lemma \ref{L:dk}, one can choose small constants $\gamma_j$,
$j=1,\dots,q$ such that $\la_{1,\gamma_j}=2\pi\alpha_j$ and $\{1,
\alpha_1,\dots,\alpha_q\}$ are rationally independent, where
$\la_{1,\gamma}$ stands for the first eigenvalue of the matrix
$\La_\gamma=\La+\gamma B$. Then using the feedback (\ref{E:con}), we
construct a time $l\ge1$ and  controls $v_j\in L^2([0,l],\R)$ such that
\begin{align}
&\dist(\RR_l(
x_j,\gamma_j+v_j),\CC_{\gamma_j})\le\frac{\delta}{2},\label{E:bnb}
\end{align}
where $\CC_{\gamma}=\{e_{1,\gamma}e^{it}:t\in\R\}$ and
$e_{1,\gamma}$ is the eigenvector corresponding to $\la_{1,\gamma}$.
Let $|\gamma_j|$ be so small that
$\|e_1-e_{1,\gamma_j}\|\le\frac{\delta}{4}$. From (\ref{E:bnb}) and
 the fact that $\{1,\alpha_1,\dots,\alpha_q\}$ are rationally
independent it follows the existence of an integer $l'\ge1$
 such that $e^{-i\La_{\gamma_j} l'}\RR_k(
z_j,\gamma_j+v_j)\in B(e_{1,\gamma_j},\frac{\delta}{4})$.
 Thus we
have constructed controls $u_j\in L^2([0,k],\R)$, $k:=l+l'$ such
that $\RR_k(x_j,u_j)\in B(e_{1},\frac{\delta}{2})$. As in the case
$\e=0$ the distance between two solutions is constant and $S\subset\cup_{j=1}^q B(x_j,\frac{\delta}{2})$, for any
$z\in S$ there is an integer $ j\in [1,q]$ such that
$\RR_k(z,u_j)\in B(e_{1},\delta)$. Clearly, as in  Step 3 in the
proof of Lemma \ref{L:dk}, we can suppose that the controls $u_j$
satisfy assertion  (ii).

\vspace{6pt} \textbf{Step 2.} Let us prove the lemma in the general
case. Let $z_0\in S$ and let $u\in L^2([0,k],\R)$ be the control
constructed in  Step 1. It is easy to see that
\begin{equation}\label{E:cgt}
\lim_{(z,\e)\rightarrow(z_0,0)
}
\RR_k^\e(z,u)=\RR_k(z_0,u).
\end{equation}
Thus there is a constant $\e(z_0)>0$ such that $\RR_k^\e(z,u)\in
 B(e_1,\delta)$ for all $|\e|<\e_0(z_0)$ and $z\in S\cap
B(z_0,\e_0(z_0))$. From the compactness of $S$ it follows  that
there is a uniform constant $\e_0>0$ such that  assertions (i) and
(ii) of Theorem~\ref{L:taza} are satisfied.
\end{proof}

\section{Proof of Lemmas \ref{L:xt} and \ref{L:tau}}\label{S:le1}
\begin{proof}[Proof of Lemma \ref{L:xt}]

\vspace{6pt} \textbf{Step 1.} To prove   assertion (i), let us
consider the mapping
\begin{align}
\RR_1^\cdot(\cdot,\cdot): (-\e_0,\e_0)\times S\cap B(e_1,\delta_0)\times X 
&\rightarrow S,\nonumber\\
(\e,z_0,u)&\rightarrow \RR_1^\e(z_0,u),\nonumber
\end{align}where
$\RR_t^\e$ is the resolving operator of problem (\ref{E:hav4}),
(\ref{E:sp4}), $X\subset L^2([0,1], \R)$ is a closed subspace and
$\e_0>0$ and $\delta_0>0$ are constants. We are going to show that
the measure $\DD(\eta_1)$ and the function
$\RR_1^\cdot(\cdot,\cdot)$
 satisfy the conditions of Theorem 2.2 in \cite{AKSS} for an
appropriate choice of $X$ and the constants $\e_0$ and $\delta_0$.
For the reader's convenience, we recall the theorem in Section
\ref{S:verg} (see Theorem \ref{T:AKSS}).

Clearly, $\RR_1^\cdot(\cdot,\cdot)$ is a  continuous function,
$\RR_1^\e(z_0,\cdot)$ is analytic for any $|\e|\le\e_0$ and $z_0\in
S\cap B(e_1,\delta_0)$, and $D_u\RR_1^\cdot(\cdot,\cdot)$ is
continuous.

Using Lemma \ref{L:1}, we see that for some positive constant $\nu$
the interior of the set $\RR_1^0(e_1,B_{E_n}(0,\nu))$ is non-empty.
Denote by $P_N$ the orthogonal projection onto the space
$span\{g_1,\dots,g_N\}$ in $L^2([0,1],\R)$. The continuity of $\RR_1^\cdot(\cdot,\cdot)$
implies that for any $\eta>0$ there is an integer $N\ge 1$ and
positive constants $\delta_0$ and $\e_0$ such that
$$
\|\RR_1^\e(z,P_Nu)-\RR_1^0(e_1,u)\|<\eta
$$for all $|\e|\le\e_0, z\in S\cap B(e_1,\delta_0)$ and $u\in
B_{E_n}(0,\nu)$. A standard degree theory argument implies  that the
interior of the set $\RR_1^\e(z,P_N(B_{E_n}(0,\nu)))$ is non-empty
for any $|\e|\le \e_0$ and $z\in S\cap B(e_1,\delta_0)$. Clearly, if
(\ref{E:pp}) holds for the integer $N$, then
  conditions of Theorem \ref{T:AKSS} are
 satisfied for $X=\overline{ span\{e_j:b_j\neq0,\, j\ge1\}}$. Thus the function
\begin{align}
{\RR_1^\cdot}_*(\cdot,\DD(\eta_1)): (-\e_0,\e_0)\times S\cap
B(e_1,\delta)&\rightarrow \PP(S),\nonumber
\end{align} is continuous, where ${\RR_1^\e}_*(z,\DD(\eta_1))$
stands for the image of the measure $\DD(\eta_1)$ under the mapping
${\RR_1^\e}(z,\cdot)$ and
 $ \PP(S)$ is endowed with the total
variation norm. This completes the  proof of assertion (i).

\vspace{6pt} \textbf{Step 2.} To prove   assertion (ii), we apply
Theorem \ref{T:AKSS} to the mapping
\begin{align}
\RR_{k+1}^\cdot(\cdot,\cdot): (-\e_0,\e_0)\times S\times X^{k+1}
&\rightarrow S,\nonumber\\
(\e,z,u)&\rightarrow \RR_{k+1}^\e(z,u),\nonumber
\end{align}where $k$ is the integer in Theorem \ref{L:taza}, $X$ is defined in
Step 1 and $X^{k+1}$ is the set of functions $u\in L^2([0,k+1],\R)$
such that $u(j+\cdot)\big|_{[0,1]}\in X$, $j=0,\dots,k$. Using
Theorem~\ref{L:taza} and the arguments of Step 1, one can show that
for sufficiently small $\e_0>0$ and for any $z\in S$ there is a ball
$B_z$ in a finite-dimensional subset of $X^{k+1}$ such that
$\RR_{k+1}^\e(z,B_z)$ has a non-empty interior for all $|\e|<\e_0$.
Clearly, the other conditions of Theorem \ref{T:AKSS} also hold.
Thus the image of the measure $\bigotimes_{j=1}^{k+1}\DD(\eta_j)$
under the mapping $\RR_{k+1}^\e(z,\cdot)$ (which is equal to
$\DD(\UU_{k+1}^\e(z))$) is absolutely continuous with respect to the
Riemannian volume $m$ on $S$.
\end{proof}

\vspace{6pt}
\begin{proof}[Proof of Lemma \ref{L:tau}]
\textbf{Step 1.} We write $\tau$ instead of $\tau_{\delta,\e}$.
 Let us show that for any $\delta>0$ there is a constant $\e_\delta>0$ such that\begin{equation}\label{E:hay}
\pP_z\{\tau <+\infty\}=1\,\,\,\,\, \text{for all $|\e|<\e_\delta$
and $z\in S$}.
\end{equation}
Using Theorem \ref{L:taza} and  Condition \ref{C:c}, one can show
that for any $\delta>0$ and $z_0\in S$ there is a time
$k=k(\delta)\ge 1$, a constant $\e_\delta=\e_\delta(z_0)>0$ and
a neighborhood $O=O(z_0)$ of $z_0$ such that
\begin{equation}
 \sup_{{(z,\e)\in O\times(-\e_\delta,\e_\delta)}}\pP_z\{\tau
>k\}<1.\nonumber
\end{equation}
From the compactness of $S$ it follows that there is a  constant
$\e_\delta>0$ such that
\begin{equation}\label{E:vah}
a:=\sup_{{(z,\e)\in S\times(-\e_\delta ,\e_\delta )}} \pP_z\{\tau
>k\}<1.
\end{equation}
Using the Markov property  and (\ref{E:vah}), we obtain
$$
\pP_z\{\tau >nk\}=\E_z(I_{\{\tau >(n-1
)k\}}\pP_{\UU_{(n-1)k}^\e(\cdot)}\{\tau >k\})\le a\pP_z\{\tau
>(n-1)k\}.
$$ Thus
\begin{equation}\label{E:vaha}
\pP_z\{\tau >nk\}\le a^n.
\end{equation}
This proves (\ref{E:hay}).

\vspace{6pt} \textbf{Step 2.} Using (\ref{E:hay}) and
(\ref{E:vaha}), we obtain for sufficiently small $\alpha>0$
\begin{align}
\E_z e^{\alpha \tau}&\le1+ \sum_{n=0}^\infty \E_z ( e^{\alpha
\tau}I_{\{nk<\tau\le(n+1)k\}})\le1+ \sum_{n=0}^\infty
e^{\alpha(n+1)k}\pP_z\{\tau>nk\}\nonumber\\&\le1+ \sum_{n=0}^\infty
e^{\alpha(n+1)k} a^n=1+\frac{e^{\alpha
    k}}{1-e^{\alpha k}a}.\nonumber
\end{align}\end{proof}
\begin{remark}\label{R:norr}An estimate similar to (\ref{E:tau}) holds for the Markov chain
$(y_n,y_n')$ constructed in  Step 2 of the proof of Theorem
\ref{T:him}. Namely, let $T$ be the stopping time defined by
(\ref{E:nort}). Let us show that (\ref{E:Tau}) holds.

Indeed, it follows from the above proof that inequality (\ref{E:Tau})
will be established if we show that for any $\delta>0$ and
$z_0,z_0'\in S$ there is a time $l=l(\delta,z_0,z_0')\ge 1$, a
constant $\e_\delta=\e_\delta(z_0,z_0')>0$ and a neighborhood
$O=O(z_0,z_0')$ of the point $(z_0,z_0')$ such that
\begin{equation}\label{E:arat}
  \sup_{{(z,z',\e)\in O\times(-\e_\delta,\e_\delta)}}\pP_{(z,z')}\{T
>l\}<1.
\end{equation}
 The case $z_0=z_0'$ follows from the
definition of the sequence $(y_n,y_n')$ and (\ref{E:vah}), and the
case $z_0,z_0'\in B(e_1,\de)$ is clear. Thus it suffices to prove
(\ref{E:arat}) in the case $z_0\neq z_0'$ and $z_0\notin
B(e_1,\de)$. Let $(\Omega,\FF,\pP)$ be the underlying probability
space. Define the event
$$
\Omega_1:=\{\omega\in\Omega:y_n=y_n'\,\,\,\text{for some $n=1,\ldots,k$}\}\in\FF,
$$ where $k\ge1$ is the integer in Theorem \ref{L:taza}. Let
 $\Omega_2:=\Omega\backslash \Omega_1$. It follows from the definition
 of  $(y_n,y_n')$ that
$$
y_k=y_k'\,\, \,\text{for any $\omega\in\Omega_1$}.
$$Using again the  definition of $(y_n,y_n')$ and
 (\ref{E:vah}), we get
\begin{equation}\label{E:mpm}
\sup_{{(z,z',\e)\in O'\times(-\e_\delta' ,\e_\delta' )}} \pP_{(z,z')}\{\tau
>2k|\Omega_1\}<1,
\end{equation}where $\e_\delta'=\e_\delta'(z_0,z_0')>0$ and
$O'=O'(z_0,z_0')$ is a neighborhood of the point $(z_0,z_0')$. On the
 other hand, by  Theorem \ref{L:taza}, Condition \ref{C:c} and the
 construction of $(y_n,y_n')$, we have
\begin{equation}\label{E:mqm}
\sup_{{(z,z',\e)\in O''\times(-\e_\delta'' ,\e_\delta'' )}} \pP_{(z,z')}\{\tau
>k|\Omega_2\}<1
\end{equation}for some  $\e_\delta''=\e_\delta''(z_0,z_0')>0$ and
$O''=O''(z_0,z_0')$. Combining (\ref{E:mpm}) and (\ref{E:mqm}), we get (\ref{E:arat}).
\end{remark}

\section{Application}\label{S:App}

In this section, we apply Theorem \ref{T:him} to the Galerkin
approximations of the Schr\"odinger equation. For any integer
$k\ge2$, denote by $Q_k$ the vector space of all polynomials of
degree $k$  with real coefficients. Let $\la$ be the Lebesgue
measure on $Q_k$.

Consider the problem
\begin{align}
i\frac{\partial z}{\partial t}  &=-z''+\beta(t) V(x)z+\e|z|^\sigma z,\,\,\,\,x\in(0,1),\label{E:hhav}\\
z(t,0)&=z(t,1)=0,\,\,\,\,\label{E:eep}\\
z(0,x)&=z_0(x),\label{E:ssp}
\end{align}
where $\sigma>0$ and $\e$ are constants, $\beta(t)$ is a random
process of the form (\ref{E:pu})  and $V\in Q_k$. Denote by
$\{e_j\}_{j\in \N}$ the set of normalized eigenfunctions
of the Dirichlet Laplacian. 
For any $n\in\N$, let
$H_n:=span\{e_1,\ldots,e_n\}$ and let $P_n$ be the orthogonal
projection onto $H_n$ in $L^2([0,1])$. The Galerkin approximation of
order $n$ of (\ref{E:hhav}) has the form
\begin{equation}\label{E:Gal}
i\frac{\partial z}{\partial t}  =-z''+\beta(t)P_n(V(x)z)+\e
P_n(|z|^\sigma z).
\end{equation}Clearly, (\ref{E:Gal}) can be rewritten in the form
(\ref{E:hav}). Denote by $\langle\cdot,\cdot\rangle$ the scalar product in
$L^2([0,1])$ and by $S$ the unit sphere in $H_n$.
\begin{theorem}\label{T:hhim}
 Suppose that Condition \ref{C:c} is  satisfied. Then for $\la$-almost all $V\in Q_k$  there is an
 integer $N\ge1$ and a constant $\e_0>0$ such that, if (\ref{E:pp}) holds, then problem
 (\ref{E:Gal}), (\ref{E:pu}) has a unique stationary measure $\mu \in
 \PP(S)$ for $|\e|<\e_0$. Moreover, $\mu$ is absolutely continuous with respect to the Riemannian volume
 on $S$ and for any initial  measure $\nu\in\PP(S)$
 inequality (\ref{E:hima}) holds.
\end{theorem}
\begin{proof}
It suffices to note that $\langle x^2e_1,e_j\rangle\neq0$ for
$j=1,\ldots,n$, and thus the set of polynomials $V\in Q_k$ with
$\langle Ve_1,e_i\rangle=0$ for some $i=1,\ldots,n$ has a zero
$\la$-measure. It remains to apply Theorem \ref{T:him}.
\end{proof}

\section{Appendix}\label{S:verg}
Here we recall a result on the finite-dimensional transformations of
measures. Let $X$ be a separable Hilbert space with the norm
$\|\cdot\|_X$. We deal with measures $\mu\in\PP(X)$ satisfying the
following condition.
\begin{condition}\label{C:C3}
 The measure $\mu\in\PP(X)$ has a finite second
  moment
$$
\int_X\|x\|_X ^2\mu(\dd x)<\infty.
$$
Moreover, there is an orthonormal basis $\{g_j\}\subset X$ such that
$$
\mu=\bigotimes_{j=1}^\infty\mu_j,
$$ where $\mu_j$ is the projection of $\mu$ to the space $X_j$ generated by
$g_j$ and $\otimes$ denotes the tensor product of measures. Finally,
for any $j\in \N$, the   measure $\mu_j$ possesses a continuous
density with respect to the Lebesgue measure on $X_j$.
\end{condition}
\begin{example}
Let $\eta$ be a random variable satisfying Condition \ref{C:c}. Then
choosing  $X:=\overline{ span\{e_j:b_j\neq0,\, j\ge1\}}$, it is easy
to see that Condition \ref{C:C3} is satisfied for the measure
$\DD(\eta)\in \PP(X)$.
\end{example}

Let $H$ be a metric space and $M$ be a finite-dimensional  analytic
Riemannian manifold.
\begin{theorem}\label{T:AKSS}
Let $f:H\times X\rightarrow M$ be a continuous function such that
$f(u,\cdot)$ is analytic for any $u\in H$ and the derivative $D_x
f(u,x)$ is continuous with respect to $(u,x)$. Suppose that, for any
$u\in H$, there is a ball $B_u$ in a finite-dimensional subspace
$X_u\subset X$ such that the interior of the set $f(u,B_u)$ is
non-empty. Then for any measure $\mu\in\PP(X)$ satisfying Condition
\ref{C:C3}, we have:
\begin{itemize}
\item[(i)]For any $u_0\in H$, the measure $f_*(u_0, \mu)$ is absolutely
  continuous with respect to the Riemannian volume on $M$, where $f_*(u_0, \mu)$
   is the image of the measure $\mu$ under the mapping $f(u_0,\cdot):X\rightarrow M$.
\item[(ii)] The function $f_*(u_0, \mu)$ from $ H$ to the space $\PP(M)$ endowed with the total
variation norm is continuous.
\end{itemize}
\end{theorem}
See \cite{AKSS} for the proof of this theorem in the case of
finite-dimensional vector space $M$. 
The result in the  case of a Riemannian manifold is deduced from the
case of finite-dimensional vector space.

Note that the main result of the paper could be stated under
 more general assumptions over the distribution of the
random variable $\eta_1$ that are adapted to a measure
transformation theorem from \cite{ABP}. Our choice is explained by
the simplicity of the conditions of Theorem \ref{T:AKSS}.

\end{document}